\documentclass[letter,twocolumn]{jpsj2} 
\title{Complete Set of Polarization Transfer Observables \\for 
the \mbox{\boldmath $^{12}{\rm C}(p,n)$} Reaction at 296~MeV and 
\mbox{\boldmath $0^{\circ}$}}

\author{Masanori \textsc{Dozono}\thanks{E-mail address: dozono@kutl.kyushu-u.ac.jp}, 
Tomotsugu \textsc{Wakasa}, Ema \textsc{Ihara}, 
Shun \textsc{Asaji}, 
Kunihiro \textsc{Fujita}$^{1}$, 
Kichiji \textsc{Hatanaka}$^{1}$, 
Takashi \textsc{Ishida}$^{2}$, 
Takaaki \textsc{Kaneda}$^{1}$, 
Hiroaki \textsc{Matsubara}$^{1}$, 
Yuji \textsc{Nagasue}, 
Tetsuo \textsc{Noro}, 
Yasuhiro \textsc{Sakemi}$^{3}$, 
Yohei \textsc{Shimizu}$^{1}$, 
Hidemitsu \textsc{Takeda}, 
Yuji \textsc{Tameshige}$^{1}$, 
Atsushi \textsc{Tamii}$^{1}$ and 
Yukiko \textsc{Yamada}
}

\inst{Department of Physics, Kyushu University, Fukuoka 812-8581 \\
$^{1}$Research Center for Nuclear Physics, Osaka University, Osaka
567-0047\\
$^{2}$Laboratory of Nuclear Science, Tohoku University, Sendai 982-0826\\
$^{3}$Cyclotron and Radioisotope Center, Tohoku University, Sendai 980-8578}

\abst{A complete set of polarization transfer observables has been
measured for the $^{12}{\rm C}(p,n)$ reaction at $T_p=296~{\rm MeV}$ 
and $\theta_{\rm lab}=0^{\circ}$. The total spin transfer 
$\Sigma(0^{\circ})$ and the observable $f_1$ deduced from the measured
polarization transfer observables indicate that 
the spin--dipole resonance at $E_x \simeq 7~{\rm MeV}$ 
has greater $2^-$ strength than $1^-$ 
strength, which is consistent with recent experimental and theoretical 
studies. 
The results also indicate a predominance of the spin-flip and 
unnatural-parity transition strength in the continuum. 
The exchange tensor interaction at a large momentum transfer 
of $Q \simeq 3.6~{\rm fm}^{-1}$ is discussed. }

\kword{complete set of polarization transfer observables, 
spin--dipole resonance, exchange tensor interaction}

\begin{document}
\maketitle

The charge exchange reaction at intermediate energies 
($T \gtrsim 100~{\rm MeV/A}$) is one of the best probes to study 
spin--isospin excitations in nuclei, such as 
spin--dipole (SD) excitations characterized by 
$\Delta L = 1,\ \Delta S = 1,\ {\rm and}\ 
\Delta J^{\pi} = 0^-,\ 1^-,\ {\rm and}\ 2^-$. 
In previous $(p,n)$ and $(n,p)$ experiments on 
$^{12}{\rm C}$,~\cite{phys.rev.C48_1158,phys.rev.C54_237} 
spin--dipole resonances (SDRs) were found 
at $E_x \simeq 4$ and 7~MeV. 
Analysis of the angular distributions of
the SDRs at $E_x \simeq 4\ {\rm and}\ 7~{\rm MeV}$ 
indicate that they consist of mainly $2^-$ and $1^-$ components, 
respectively. 
However, a recent $^{12}{\rm C}(\vec{d},{}^2{\rm He})^{12}{\rm B}$ 
experiment~\cite{phys.rev.C66_054602} 
suggested that 
the SDR at $E_x \simeq 7~{\rm MeV}$ in $^{12}{\rm B}$ 
has more $2^-$ components than $1^-$ components. 
This suggestion is supported by a 
$^{12}{\rm C}(^{12}{\rm C},{}^{12}{\rm N})^{12}{\rm B}$
experiment~\cite{nucl.phys.A577_93c} and by 
theoretical calculations including tensor correlations.~\cite{nucl.phys.A637_547}
Thus the spin-parity assignment of the SDR at $E_x \simeq 7~{\rm MeV}$ 
for the $A=12$ system is still controversial. 

A complete set of polarization transfer (PT) observables at $0^{\circ}$ 
is a powerful tool for investigating the spin-parity $J^{\pi}$ of 
an excited state. 
The total spin transfer $\Sigma(0^{\circ})$ deduced from such a set gives 
information on the transferred spin $\Delta S$, which is independent of 
theoretical models.~\cite{prog.theor.phys.86_1129} 
Furthermore, information can be obtained on the parity from the 
observable $f_1$.~\cite{j.phys.soc.jpn.73_1611} 
On the other hand, each PT observable is sensitive to 
the effective nucleon--nucleon ($NN$) interaction. 
The PT observables for $\Delta J^{\pi} = 1^+$ transitions 
have been used to study the exchange tensor interaction at large momentum 
transfers.~\cite{phys.rev.lett.71_684,phys.rev.C51_R2871} 

In this Letter, we present measurements of a complete set of PT
observables for the $^{12}{\rm C}(p,n)$ reaction at $T_p=296~{\rm MeV}$ 
and $\theta_{\rm lab}=0^{\circ}$. We have deduced the total spin transfer 
$\Sigma$ and the observable $f_1$ using the measured PT observables in 
order to investigate the spin-parity structure 
in both the SDR and continuum regions. 
We also compare the PT observables for 
the $^{12}{\rm C}(p,n)^{12}{\rm N}({\rm g.s.};1^+)$ reaction 
with distorted-wave impulse approximation (DWIA) calculations 
employing the effective $NN$ interaction 
in order to assess the effective tensor interaction 
at a large exchange momentum transfer of $Q \simeq 3.6~{\rm fm}^{-1}$. 

Measurements were carried out at 
the neutron time-of-flight facility~\cite{nimA369_120}
at the Research Center for Nuclear Physics (RCNP), Osaka University. 
The proton beam energy was 296~MeV and 
the typical current and polarization were 500~nA and 0.70, respectively. 
The neutron energy and polarization were measured by 
the neutron detector/polarimeter NPOL3.~\cite{nimA547_569} 
We used a natural carbon $(98.9\%\ ^{12}{\rm C})$ 
target with a thickness of 89~mg/cm$^2$. 
The measured cross sections were normalized to the 
$0^{\circ}\ ^{7}{\rm Li}(p,n)^{7}{\rm Be}({\rm g.s.}+ 0.43~{\rm MeV})$ 
reaction, which has a center of mass (c.m.) cross section of 
$\sigma_{\rm c.m.}(0^{\circ})=27.0 \pm 0.8~{\rm mb/sr}$ 
at this incident energy.~\cite{phys.rev.C41_2548} 
The systematic uncertainties of the data were estimated to be 4--6\%. 

Asymmetries of the $^{1}{\rm H}(\vec{n},p)n$ and 
$^{12}{\rm C}(\vec{n},p){\rm X}$ reactions in NPOL3 were used 
to deduce the neutron polarization. 
The effective analyzing power $A_{y{\rm ;eff}}$ of NPOL3 was 
calibrated by using polarized neutrons from the 
$^{12}{\rm C}(\vec{p},\vec{n})^{12}{\rm N}({\rm g.s.;}1^+)$ 
reaction at 296~MeV and $0^{\circ}$.
A detailed description of the calibration can be found in 
Ref.~\citen{nimA547_569}. 
The resulting $A_{y{\rm ;eff}}$ was 
$0.151 \pm 0.007 \pm 0.004$, where the first and second uncertainties 
are statistical and systematic, respectively. 

\begin{figure}
\begin{center}
\includegraphics[width=0.82\hsize]{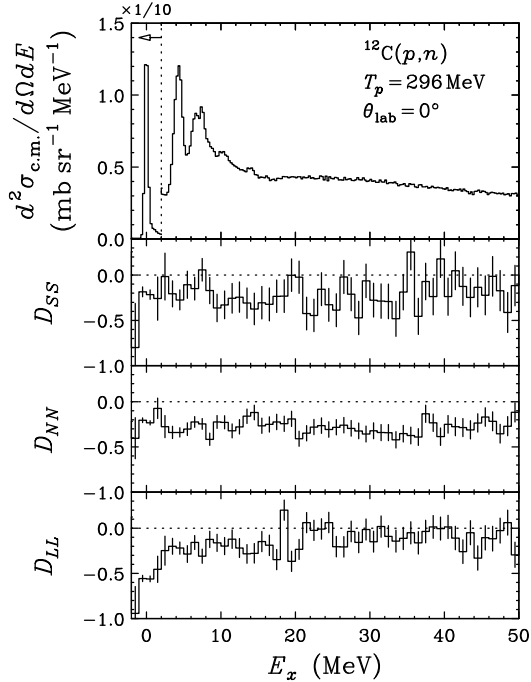}
\caption{Double differential cross section (top panel) and a complete 
set of polarization transfer observables (bottom three panels) for the 
$^{12}{\rm C}(p,n)$ reaction at $T_p=296~{\rm MeV}$ and 
$\theta_{\rm lab}=0^{\circ}$. The error bars represent statistical 
uncertainties only.}
\label{fig_CS_Dii}
\end{center}
\end{figure}

Figure~\ref{fig_CS_Dii} shows the double differential cross section and 
a complete set of PT observables $D_{ii}\ (i=S,\ N,\ {\rm and}\ L)$ 
at $0^{\circ}$ 
as a function of excitation energy $E_x$. 
The laboratory coordinates at $0^{\circ}$ are defined so that 
the normal ($\hat{\mbox{\boldmath $N$}}$) direction is the same as 
$\hat{\mbox{\boldmath $N$}}$ at finite angles 
(normal to the reaction plane), 
the longitudinal ($\hat{\mbox{\boldmath $L$}}$) 
direction is along the momentum transfer, and the sideways 
($\hat{\mbox{\boldmath $S$}}$) direction is given by 
$\hat{\mbox{\boldmath $S$}}=\hat{\mbox{\boldmath $N$}} \times 
\hat{\mbox{\boldmath $L$}}$. 
The data of the cross section in Fig.~\ref{fig_CS_Dii} have been 
sorted into 0.25-MeV bins, while the data of 
$D_{ii}(0^{\circ})$ have been sorted into 1-MeV bins to reduce
statistical fluctuations. 
A high energy resolution of 500~keV full width at half maximum 
(FWHM) was realized by NPOL3, which enabled us to observe clearly 
two SDR peaks at $E_x \simeq 4\ {\rm and}\ 7~{\rm MeV}$. 
It should be noted that the $D_{NN}(0^{\circ})$ value should be equal to 
the corresponding $D_{SS}(0^{\circ})$ value 
because the $\hat{\mbox{\boldmath $N$}}$ direction is identical 
to the $\hat{\mbox{\boldmath $S$}}$ direction at $0^{\circ}$. 
The experimental $D_{NN}(0^{\circ})$ and $D_{SS}(0^{\circ})$ values 
are consistent with each other within statistical uncertainties 
over the entire range of $E_x$, demonstrating the reliability of 
our measurements. 

\begin{figure}
\begin{center}
\includegraphics[width=0.77\hsize]{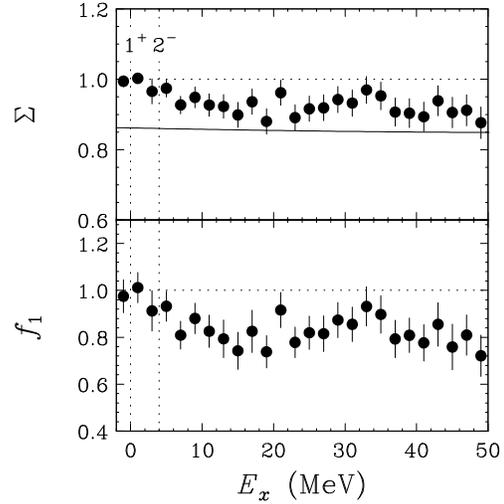}
\caption{Total spin transfer $\Sigma$ (top panel) and 
observable $f_1$ (bottom panel) for the 
$^{12}{\rm C}(p,n)$ reaction at $T_p=296~{\rm MeV}$ and 
$\theta_{\rm lab}=0^{\circ}$. 
The error bars represent statistical uncertainties only.
The solid line shows the values of $\Sigma$ for free 
$NN$ scattering.}
\label{fig_Sigma_f1}
\end{center}
\end{figure}

Figure~\ref{fig_Sigma_f1} shows the total spin transfer 
$\Sigma (0^{\circ})$ and the observable $f_1$ defined 
as~\cite{prog.theor.phys.86_1129,j.phys.soc.jpn.73_1611} 
\begin{equation}
\begin{array}{lll}
\Sigma(0^{\circ})& = & 
\displaystyle{\frac{3-[2D_{NN}(0^{\circ})+D_{LL}(0^{\circ})]}{4}},\\[15pt]
f_1 & = &
\displaystyle{\frac{1-2D_{NN}(0^{\circ})+D_{LL}(0^{\circ})}
{2[1+D_{LL}(0^{\circ})]}},
\end{array}
\end{equation}
as a function of excitation energy $E_x$. 
The $\Sigma(0^{\circ})$ value is either 0 or 1 
depending on whether $\Delta S = 0$ or $\Delta S = 1$, 
which is independent of theoretical models.~\cite{prog.theor.phys.86_1129} 
The $f_1$ value is either 0 or 1 depending on the natural-parity
or unnatural-parity transition if a single $\Delta J^{\pi}$ 
transition is dominant.~\cite{j.phys.soc.jpn.73_1611} 
The $\Sigma(0^{\circ})$ and $f_1$ values of 
the spin-flip unnatural-parity $1^+$ and $2^-$ states 
at $E_x=0$ and 4~MeV, respectively, are almost unity, 
which is consistent with theoretical predictions. 
The continuum $\Sigma(0^{\circ})$ values are almost 
independent of $E_x$ and take values larger than $0.88$ up to 
$E_x = 50~{\rm MeV}$, indicating the predominance of the spin-flip strength. 
The solid line in the top panel of 
Fig.~\ref{fig_Sigma_f1} 
represents the free $NN$ values of $\Sigma(0^{\circ})$ for the 
corresponding kinematical condition.~\cite{said} 
Enhancement of $\Sigma(0^{\circ})$ relative to 
the free $NN$ values means enhancement of  
the $\Delta S=1$ response relative to the $\Delta S=0$ response 
in nuclei at small momentum transfers, which 
is consistent with previous studies 
of $(p,p')$ scattering.~\cite{phys.rev.lett.58_2404,phys.lett.B237_337} 
The large values of $f_1 \ge 0.72$ up to $E_x=50~{\rm MeV}$ indicate 
a predominance of the unnatural-parity transition strength in the 
continuum, consistent with the $^{90}{\rm Zr}(p,n)$ result 
at 295~MeV.~\cite{j.phys.soc.jpn.73_1611} 

\begin{figure}
\begin{center}
\includegraphics[width=0.80\hsize]{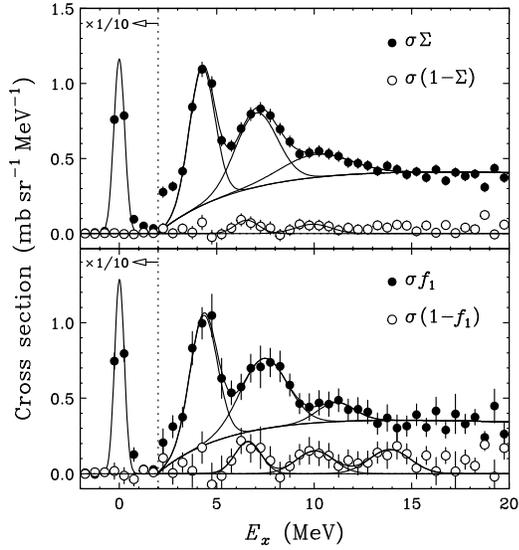}
\caption{Cross sections separated by $\Sigma$ (top panel) and 
$f_1$ (bottom panel) for the 
$^{12}{\rm C}(p,n)$ reaction at $T_p=296~{\rm MeV}$ and 
$\theta_{\rm lab}=0^{\circ}$. 
The solid lines show peak fitting of the spectra 
with Gaussian peaks and a continuum. }
\label{fig_separated_CS}
\end{center}
\end{figure}

\begin{figure}
\begin{center}
\includegraphics[width=0.87\hsize]{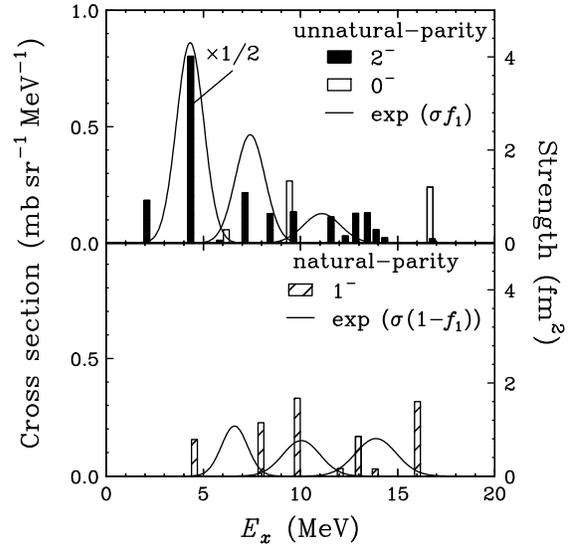}
\caption{SD strengths for unnatural-parity (top panel) 
and natural-parity (bottom panel) 
taken from Ref.~\citen{nucl.phys.A637_547}. 
The solid lines represent peaks obtained by 
fitting $\sigma f_1$ (top panel) and 
$\sigma (1-f_1)$ (bottom panel) spectra. }

\label{fig_SD_strength}
\end{center}
\end{figure}

The top panel of Fig.~\ref{fig_separated_CS} shows the spin-flip 
($\sigma \Sigma$) and non-spin-flip ($\sigma (1-\Sigma)$) cross
sections as filled and open circles, respectively, 
as functions of $E_x$. 
The bottom panel shows the unnatural-parity dominant ($\sigma f_1$)
and natural-parity dominant ($\sigma(1-f_1)$) components of the 
cross section as filled and open circles, respectively. 
The solid lines are the results of peak fitting of the spectra 
with Gaussian peaks and a continuum. 
The continuum was assumed to be the quasi-free scattering contribution, 
and its shape was given by the formula 
given in Ref.~\citen{phys.rev.C34_1822}. 
It should be noted that the spin-flip unnatural-parity 
$1^{+}$ and $2^{-}$ states at $E_x=0$ and 4~MeV, respectively, 
form peaks only in the $\sigma \Sigma$ and $\sigma f_1$ spectra. 
It is found that the prominent peak at $E_x \simeq 7~{\rm MeV}$ is the 
spin-flip unnatural-parity component with a $J^{\pi}$ value 
estimated to be $2^{-}$ because the $D_{ii}(0^{\circ})$ values 
are consistent with the theoretical prediction 
for $J^{\pi}=2^-$.~\cite{phys.rev.C26_727} 
In the $\sigma(1-f_1)$ spectrum, 
possible evidence for SD $1^-$ peaks is seen 
at $E_x \simeq 7,\ 10,\ {\rm and}\ 14~{\rm MeV}$. 
The top and bottom panels of Fig.~\ref{fig_SD_strength} 
show theoretical calculations for the unnatural-parity and natural-parity 
SD strengths, respectively.~\cite{nucl.phys.A637_547} 
Experimentally extracted peaks in the $\sigma f_1$ and 
$\sigma (1-f_1)$ spectra are also shown. 
Concentration of the SD $2^-$ strength at three peaks at 
$E_x \simeq 4,\ 8,\ {\rm and}\ 13~{\rm MeV}$ has been predicted. 
Our data agree with this prediction qualitatively, 
but give slightly different excitation energies of 
$E_x \simeq 4,\ 7,\ {\rm and}\ 11~{\rm MeV}$. 
On the other hand, 
the SD $1^-$ strength has been predicted to be 
quenched and fragmented due to tensor correlations.~\cite{nucl.phys.A637_547}  
The experimental results are spread over a wide region of 
$E_x \simeq 5$--$16~{\rm MeV}$ and exhibit similar cross sections, 
which supports fragmentation of the SD $1^-$ strength. 

Effective tensor interactions at $q \simeq1$--3~${\rm fm}^{-1}$ 
have mainly been studied using high spin stretched 
states.~\cite{edward,phys.rev.C45_1098} 
The present $D_{ii}(0^{\circ})$ data can give information on 
the exchange tensor interaction at an extremely large exchange 
momentum transfer of $Q \simeq 3.6~{\rm fm}^{-1}$. 
In the Kerman--McNanus--Thaler (KMT) 
representation~\cite{ann.phys.n.y.8_551}, 
the $NN$ scattering amplitude is represented as 
\begin{equation}
\begin{array}{rl}
M(q) = & \hspace{-10pt}A + \frac{1}{3} (B+E+F) 
\mbox{\boldmath $\sigma$}_1 \cdot \mbox{\boldmath $\sigma$}_2 
+ C (\mbox{\boldmath $\sigma$}_1 + \mbox{\boldmath $\sigma$}_2)
 \cdot \mbox{\boldmath $\hat{n}$}\\
& \hspace{-10pt} + \frac{1}{3}(E-B) S_{12}(\mbox{\boldmath $\hat{q}$})
+ \frac{1}{3}(F-B) S_{12}(\mbox{\boldmath $\hat{Q}$}),
\end{array}
\end{equation}
where $S_{12}$ is the tensor operator, 
$\mbox{\boldmath $\hat{q}$}$ and $\mbox{\boldmath $\hat{Q}$}$ 
are direct and exchange momentum transfers, respectively, and 
$\mbox{\boldmath $\hat{n}$}=\mbox{\boldmath $\hat{Q}$}
\times\mbox{\boldmath $\hat{q}$}$. 
In a plane-wave impulse approximation (PWIA), 
the PT observables for the Gamow--Teller (GT) transition at $0^{\circ}$ are 
simply expressed using parameters $A$--$F$ as~\cite{phys.rev.C26_727} 
\begin{equation}
\begin{array}{llcll}
D_{NN}(0^{\circ})&\hspace{-10pt}=&\hspace{-10pt}
D_{SS}(0^{\circ})
&\hspace{-10pt}=&\hspace{-10pt}\displaystyle{\frac{-F^2}{2B^2+F^2}},
\\[10pt]
D_{LL}(0^{\circ})&\hspace{-10pt}=&\hspace{-10pt}
\displaystyle{\frac{-2B^2+F^2}{2B^2+F^2}}.&&
\end{array}
\end{equation}
If there is no exchange tensor $S_{12}(\mbox{\boldmath $\hat{Q}$})$ 
interaction (i.e., $F=B$), then $D_{ii}(0^{\circ})=-1/3$.

The measured PT observables $D_{ii}(0^{\circ})$ for the GT 
$^{12}{\rm C}(\vec{p},\vec{n})^{12}{\rm N}({\rm g.s.;}1^+)$ 
transition are listed in Table~\ref{tab_Dii}, where 
the listed uncertainties are statistical only. 
The present $D_{NN}(0^{\circ})$ and $D_{SS}(0^{\circ})$
values are consistent with each other, as expected, 
and the present $D_{NN}(0^{\circ})$ value agrees with the 
previously measured $D_{NN}(0^{\circ})$ value 
at the same energy.~\cite{phys.rev.C51_R2871} 
The experimental values deviated from $-1/3$, which 
indicates that there are contributions from both the exchange tensor interaction 
at $Q \simeq 3.6~{\rm fm}^{-1}$ and nuclear distortion effects. 

In order to assess these effects quantitatively, 
we performed microscopic DWIA calculations 
using the computer code {\sc dw81}~\cite{dw81}. 
The transition amplitudes were calculated from the Cohen--Kurath wave 
functions~\cite{nucl.phys.73_1} assuming 
Woods--Saxon radial dependence.~\cite{phys.rev.C38_589} 
Distorted waves were generated using the optical model potential 
(OMP) for proton elastic scattering data on $^{12}{\rm C}$ 
at 318~MeV.~\cite{phys.rev.C48_1106} 
We used the effective $NN$ interaction parameterized by Franey and Love (FL) 
at 270 or 325~MeV.~\cite{phys.rev.C31_488}

First, we examined the sensitivity of the DWIA results to the OMPs 
by using two different parameters.~\cite{phys.rev.C48_1106,phys.rev.C31_1569} 
The OMP dependence of $D_{ii}(0^{\circ})$ was found to be less 
than 0.01. 
This insensitivity allows us to use $D_{ii}(0^{\circ})$ as 
a probe to study the effective $NN$ interaction. 
Table~\ref{tab_Dii} shows the DWIA results for $D_{ii}(0^{\circ})$ 
with the $NN$ interaction at 270 and 325~MeV. 
It is found that the $D_{ii}(0^{\circ})$ values, 
and $D_{LL}(0^{\circ})$ in particular, are 
sensitive to the choice of the $NN$ interaction. 
These differences are mainly due to the exchange tensor interaction 
$S_{12}(Q)$ at $Q \simeq 3.6~{\rm fm}^{-1}$. 
The real part of $S_{12}(Q)$ for the FL 325~MeV interaction is about 
twice as large as that for the FL 270~MeV interaction at 
$Q \simeq 3.6~{\rm fm}^{-1}$ 
(see Fig.~3 of Ref.~\citen{phys.rev.C51_R2871}). 
The experimental $D_{ii}(0^{\circ})$ values support the DWIA 
results with the FL 270~MeV interaction, which 
indicates that the exchange tensor part of the FL 270~MeV 
interaction has an appropriate strength at $Q \simeq 3.6~{\rm fm}^{-1}$. 
This conclusion has already been reported for 
$D_{NN}(0^{\circ})$ data,~\cite{phys.rev.C51_R2871} however, 
the present data make the conclusion more rigorous 
because of the high sensitivity of $D_{LL}(0^{\circ})$ 
to the exchange tensor interaction. 

\begin{table*}[t]
\newcommand{\lw}[1]{\smash{\lower2.0ex\hbox{#1}}}
\begin{center}
\renewcommand{\arraystretch}{1.2}
\begin{tabular}{ccccc}\hline \hline
& & $D_{NN}(0^{\circ})$ & $D_{SS}(0^{\circ})$ & $D_{LL}(0^{\circ})$\\ \hline
\lw{Exp.} & This work & $-0.216 \pm 0.019$ & $-0.210 \pm 0.039$ 
& $-0.554 \pm 0.023$\\
& ref.~\citen{phys.rev.C51_R2871} & $-0.215 \pm 0.019$ & -- & -- \\ \hline
\lw{DWIA} & FL 270~MeV & $-0.225$ & $-0.225$ & $-0.550$\\
& FL 325~MeV & $-0.191$ & $-0.191$ & $-0.619$\\ \hline \hline
\end{tabular}
\end{center}
\caption{PT observables $D_{ii}(0^{\circ})$ for the GT
$^{12}{\rm C}(\vec{p},\vec{n})^{12}{\rm N}({\rm g.s.;}1^+)$ transition 
at 296~MeV and $0^{\circ}$ compared with theoretical calculations.}
\label{tab_Dii}
\end{table*}

In summary, a complete set of PT observables for the 
$^{12}{\rm C}(p,n)$ reaction at $T_p=296~{\rm MeV}$ and 
$\theta_{\rm lab}=0^{\circ}$ has been measured. 
The total spin transfer $\Sigma(0^{\circ})$ and the observable $f_1$ 
are deduced in order to study the spin-parity structure in 
both the SDR and continuum regions. 
The $\Sigma(0^{\circ})$ and $f_1$ values show that 
the SDR at $E_x \simeq 7~{\rm MeV}$ has 
greater $2^-$ strength than $1^-$ strength, 
which agrees with the recent theoretical prediction.
In the continuum up to $E_x \simeq 50~{\rm MeV}$, 
a predominance of the spin-flip and 
unnatural-parity transition strength 
is also found. 
We have compared the PT observables of the 
$^{12}{\rm C}(p,n)^{12}{\rm N}({\rm g.s.;}1^+)$ reaction 
with DWIA calculations employing the FL interaction. 
The exchange tensor interaction of the FL 270~MeV interaction is found 
to be more appropriate at $Q \simeq 3.6~{\rm fm}^{-1}$ 
than that of the FL 325~MeV interaction. 
Thus a complete set of PT observables provides rigorous information 
not only on the spin-parity structure in nuclei 
but also on the effective $NN$ interaction.

\section*{Acknowledgment}
We are grateful to the RCNP cyclotron crew for providing a good quality 
beam for our experiments. We also thank H. Tanabe for his help during 
the experiments. This work was supported in part by the Grants-in-Aid 
for Scientific Research Nos. 14702005 and 16654064 of the Ministry of 
Education, Culture, Sports, Science, and Technology of Japan.

\end{document}